%% file: main.tex
\documentclass[preprint,5p,times,twocolumn]{elsarticle}
\usepackage{amsmath,amssymb,amsfonts}
\usepackage{algorithmic}

\usepackage[nolist, withpage]{acronym}
\usepackage{pifont}
\usepackage{tabularx}
\usepackage{multirow}
\usepackage{makeidx}  
\makeindex
\usepackage{pifont}
\usepackage{url}
\usepackage{subfigure}
\usepackage{color}
\usepackage{multirow}
\usepackage{listings}
\usepackage[ruled,boxed, vlined,linesnumbered,resetcount]{algorithm2e}
\usepackage{tikz}

\usepackage{afterpage}
\usepackage{placeins}

\usepackage{hyperref}
\hypersetup{
      colorlinks,%
      citecolor=blue,%
      filecolor=black,%
      linkcolor=black,%
      urlcolor=black
}

\newlength{\figsize}
\newlength{\subfigwidth}
\newlength{\subfiglabelwidth}
\input{macros}

\newcommand{\circled}[1]{\tikz[baseline=(myanchor.base)] \node[circle,fill=.,inner sep=1pt] (myanchor) {\color{-.}\bfseries\footnotesize #1};}
\journal{Computer Communications Journal, Elsevier}

\begin{document}

\begin{frontmatter}



\title{Autonomous Integration of TSN-unaware Applications with QoS Requirements in TSN Networks}

\affiliation[knchair]{organization={University~of~Tuebingen, Chair~of~Communication~Networks},
            addressline={Sand 13}, 
            city={Tuebingen},
            postcode={72076}, 
            state={Baden-Wuerttemberg},
            country={Germany}
            }
\affiliation[rhebo]{organization={Rhebo GmbH - a Landis+Gyr Company},
            addressline={Spinnereistr. 7}, 
            city={Leipzig},
            postcode={04179}, 
            state={Sachsen},
            country={Germany}
            }

\author[knchair]{Moritz Fluechter}\ead{moritz.fluechter@uni-tuebingen.de}
\author[knchair]{Steffen Lindner}\ead{steffen.lindner@uni-tuebingen.de}
\author[knchair]{Lukas Osswald}\ead{lukas.osswald@uni-tuebingen.de}
\author[rhebo]{Jérôme Arnaud}\ead{jerome.arnaud@rhebo.com}
\author[knchair]{Michael Menth}\ead{menth@uni-tuebingen.de}

\input{chapters/abstract}




\end{frontmatter}


\input{chapters/introduction}

\input{chapters/background}
\input{chapters/related_work}
\input{chapters/neural_networks}
\input{chapters/concept}
\input{chapters/periodicity}

\input{chapters/traffic_description_algorithm}
\input{chapters/qos_classification}
\input{chapters/evaluation}
\input{chapters/discussion}
\input{chapters/conclusion}

\section*{Acknowledgements}
This work has been supported by the German Federal Ministry of Education and Research (BMBF) under support code 16KIS1161 (Collaborative Project KITOS).
The authors alone are responsible for the content of the paper.

\input{acronyms.tex}

\bibliographystyle{elsarticle-num} 
\bibliography{bib/Literature}





\end{document}

%% file: macros.tex
\newcommand\fig[1]{Figure~\ref{fig:#1}}

\newcommand\sect[1]{Section~\ref{sec:#1}}
\newcommand\tabl[1]{Table~\ref{tab:#1}}

\newcommand{\floor}[1]{\left\lfloor #1 \right\rfloor}

\newcommand{\figeps}[3][]{%
 \begin{figure}[htbp]
  \begin{center}
   \leavevmode
      \parbox[t]{#1}{%
        \resizebox{#1}{!}{\includegraphics{figures/#2}}
      }
   \caption{#3\vspace{-0.2cm}}
   \label{fig:#2}
  \end{center}
 \end{figure}
}

\newenvironment{tab}[2]{%
 \begin{table}[htbp!]
  \begin{center}
  \caption{#2\label{tab:#1}}
}{%
  \end{center}
 \end{table}
}

\newcommand{\foursubfigeps}[9]{
  \begin{figure*}[htbp!]
    \leavevmode
    \begin{center}
     \subfigure[#2]{
        \label{fig:#1}
        \parbox[t]{0.22\textwidth}{%
            \resizebox{0.22\textwidth}{!}{\includegraphics{figures/#1}}
     \vspace{-1cm}
        }
     }
     \subfigure[#4]{
        \label{fig:#3}
        \parbox[t]{0.22\textwidth}{%
            \resizebox{0.22\textwidth}{!}{\includegraphics{figures/#3}}
     \vspace{-1cm}
        }
     }
     \subfigure[#6]{
        \label{fig:#5}
        \parbox[t]{0.22\textwidth}{%
            \resizebox{0.22\textwidth}{!}{\includegraphics{figures/#5}}
        }
     }
     \subfigure[#8]{
        \label{fig:#7}
        \parbox[t]{0.22\textwidth}{%
            \resizebox{0.22\textwidth}{!}{\includegraphics{figures/#7}}
        }
     }
    \end{center}
    \caption{#9}
  \end{figure*}
}

\newcommand{\foursubfigverticaleps}[9]{
  \begin{figure}[htbp!]
    \leavevmode
    \begin{center}
     \subfigure[#2]{
        \label{fig:#1}
        \parbox[t]{0.9\columnwidth}{%
            \resizebox{0.9\columnwidth}{!}{\includegraphics{figures/#1}}
        }
     }
     \end{center}
     \begin{center}
     \subfigure[#4]{
        \label{fig:#3}
        \parbox[t]{0.9\columnwidth}{%
            \resizebox{0.9\columnwidth}{!}{\includegraphics{figures/#3}}
        }
     }
     \end{center}
     \begin{center}
     \subfigure[#6]{
        \label{fig:#5}
        \parbox[t]{0.9\columnwidth}{%
            \resizebox{0.9\columnwidth}{!}{\includegraphics{figures/#5}}
        }
     }
     \end{center}
     \begin{center}
     \subfigure[#8]{
        \label{fig:#7}
        \parbox[t]{0.9\columnwidth}{%
            \resizebox{0.9\columnwidth}{!}{\includegraphics{figures/#7}}
        }
     }
    \end{center}
    \caption{#9}
  \end{figure}
}

\newboolean{makevspace}
\newcommand{\cvspace}[1]{%
    \ifthenelse
        {\boolean{makevspace}}
        {\vspace{#1}}
        {}%
    }

%% file: chapters/abstract.tex
\begin{abstract}
Modern industrial networks transport both best-effort and real-time traffic.
\acf{TSN} was introduced by the IEEE \ac{TSN} Task Group as an enhancement to Ethernet to provide high \ac{QoS} for real-time traffic.
In a \ac{TSN} network, applications signal their \ac{QoS} requirements to the network before transmitting data.
The network then allocates resources to meet these requirements.
However, \ac{TSN}-unaware applications can neither perform this registration process nor profit from \ac{TSN}'s \ac{QoS} benefits.

The contributions of this paper are twofold.
First, we introduce a novel network architecture in which an additional device autonomously signals the \ac{QoS} requirements of \ac{TSN}-unaware applications to the network.
Second, we propose a processing method to detect real-time streams in a network and extract the necessary information for the \ac{TSN} stream signaling.
It leverages a \acf{DRNN} to detect periodic traffic, extracts an accurate traffic description, and uses traffic classification to determine the source application.
As a result, our proposal allows \ac{TSN}-unaware applications to benefit from \acp{TSN} \ac{QoS} guarantees.
Our evaluations underline the effectiveness of the proposed architecture and processing method.
\end{abstract}

%% file: chapters/introduction.tex
\section{Introduction}
\label{sec:introduction}

Applications in an industrial setting, e.g., factory automation, rely on networks offering high \acf{QoS}.
\acf{TSN} is a set of standards that defines appropriate real-time features in Ethernet networks while also supporting \acf{BE} traffic. 
It has emerged as a successor to \acf{AVB} and is currently being standardized by the IEEE 802.1 TSN Task Group.
\ac{TSN} comprises protocols and concepts for resource and network management, time synchronization, traffic shaping, and reliability.
Applications with \ac{TSN} support communicate their \ac{QoS} requirements to the network before transmitting data.
This excludes legacy or generally \ac{TSN}-unaware systems that do not support \ac{TSN} signaling but still have real-time requirements.
Examples are legacy soft-realtime applications, such as control-to-control traffic of industrial \acfp{PLC}, or real-time traffic in a converged network, e.g., video streams.

In this paper, we propose an architecture that integrates streams from \ac{TSN}-unaware systems into the \ac{TSN} network.
It uses a novel network monitoring entity, called \ac{SCIP}, divided into four processing stages: stream recognition, periodicity detection, traffic description extraction, and \ac{QoS} classification.
Within the stream recognition stage, streams are monitored, and packets are recorded for processing in the subsequent stages.
In the periodicity detection stage, streams are classified as periodic or aperiodic.
Only periodic streams are further considered for an automated integration.
We leverage a \acf{DRNN} for that purpose.
Afterward, the \ac{TSN} traffic description is retrieved from the recorded packets in the traffic description extraction stage.
Finally, the \ac{QoS} requirements are determined via traffic classification.
We evaluate our concept through simulations with the discrete event simulator OMNeT++~\cite{noauthor_omnet_nodate} with the INET framework~\cite{noauthor_inet_nodate} and show the benefits for a \acf{VoIP} stream in an overloaded network.

\par \medskip
The remainder of this paper is structured as follows.
In \sect{related_work}, we review related work.
Afterward, we summarize the relevant concepts of \acf{TSN} and the \ac{TSN} stream announcement process in \sect{background}.
In \sect{neuralnetworks}, we introduce the fundamentals of \acfp{NN} and the concept of \acfp{RNN}.
We present our novel \acf{SCIP} and the different processing stages in \sect{concept}.
In \sect{periodicity}, we elaborate on our approach to detect periodic streams with a \ac{RNN} and evaluate it on an artificial dataset.
We present an algorithm to automatically extract the \ac{TSN} traffic description of an observed stream in \sect{traffic_description}.
In \sect{qos_classification}, we present the concept of deriving \ac{QoS} requirements via traffic classification.
We evaluate the presented concept through a simulation with the OMNeT++ simulator and INET framework in \sect{evaluation}.
Finally, we analyze the compatibility of the proposed architecture with existing \ac{TSN} mechanisms in \sect{discussion} and conclude the paper in \sect{conclusion}.

%% file: chapters/background.tex
\section{\acf{TSN}}\label{sec:background}
First, we introduce the fundamentals of \ac{TSN}.
We then explain how \ac{TSN} bridges determine which \ac{TSN} mechanisms to apply to a packet.
Finally, we explain the resource reservation process in a \ac{TSN} network.

\subsection{Fundamentals of \acf{TSN}}

\ac{TSN} is a set of standards that extends Ethernet with real-time capabilities.
Transmissions in a \ac{TSN} network can benefit from \ac{QoS} guarantees such as bounded delay, no congestion-based packet loss, or low jitter.
In the context of \ac{TSN}, the transmissions are called streams and the participating devices end stations.
More specifically, the source is called the talker, and the destination listener.
A \ac{TSN} stream may have multiple listeners but only one talker.
Before a talker can start the data transmission, the stream has to be admitted by the network.
During this admission control process, the talker and listener signal information about the stream, such as the transmission rate, source and destination, and \ac{QoS} requirements.
The network is configured by a central entity or through a distributed mechanism such that the previously signaled \ac{QoS} requirements are met.
Thereby, policing and shaping mechanisms are configured on the bridges along the forwarding path.
This includes a configuration of stream identification functions, as defined in IEEE Std. 802.1CB~\cite{8021cb} to associate incoming packets with the registered stream.

\subsection{Admission Control}

The IEEE Std. 802.1Qcc~\cite{8021qcc} defines three network architecture models for admission control: fully centralized, fully distributed, and centralized network/distributed user.
In this work, we focus only on the fully centralized network model illustrated in \fig{FIG1.pdf}.

\figeps[\columnwidth]{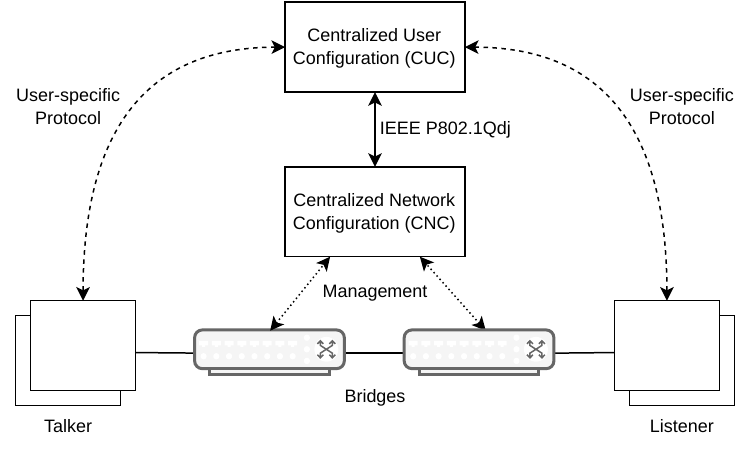}{The fully centralized configuration model is composed of end stations, bridges, one or more \acs{CUC}s, and a single \acs{CNC}. IEEE P802.1Qdj~\cite{8021qdj} defines the communication interface between CUC and CNC.}

The fully centralized network model introduces two entities for network management, the so-called \acf{CNC} and \acf{CUC}.
End stations, i.e., talkers and listeners, leverage a user-specific protocol to signal the stream properties and requirements to the \ac{CUC} before transmitting data.
The CUC collects stream properties and requirements of all end stations participating in the same stream.
Then, the \ac{CUC} initiates the resource reservation process with the \ac{CNC} using the collected data.
This communication is achieved via the interface defined in IEEE Std. P802.1Qdj~\cite{8021qdj}.
The \ac{CNC} performs admission control based on the requested resources.
It calculates the network configuration for bridges and end stations to guarantee the requested \ac{QoS}.
If the admission is successful, the \ac{CNC} configures the bridges via a network configuration protocol, e.g., NETCONF~\cite{netconf} or SNMP~\cite{snmp}.
Afterwards, it signals the result of the admission control and configuration for the end stations back to the \ac{CUC}.
This interface configuration includes the \ac{VLAN} ID and \ac{PCP} value that the end station has to set in every packet.
The \ac{CUC} forwards the admission result and the interface configuration to the end stations via the user-specific protocol.
Then, the end stations can start to transmit stream data.
This process has to be performed for every \ac{TSN} stream, but the \ac{CUC} may announce multiple streams at the same time.



%% file: chapters/related_work.tex
\section{Related Work}\label{sec:related_work}
We first summarize related work that aims to integrate legacy network traffic into TSN networks.
Then, we discuss work that focuses on detecting periodicity in time series data.
\subsection{Integrating Legacy Traffic into TSN}
Gavriluţ and Pop~\cite{gavrilut_traffic-type_2020} present an approach focusing on soft and hard real-time streams.
They propose an offline metaheuristic that assigns streams to one of three traffic classes: \acf{TT}, \acf{AVB}, and \acf{BE}.
Afterwards, the network operator has to manually configure the TSN network such that the \ac{QoS} requirements are met.
This approach requires knowledge of the network topology and manual configuration.
The LETRA tool by Mateu et al.~\cite{MaAs21} leverages a similar approach.
It uses multiple characteristics of the streams, e.g., maximal delay, period, and real-time requirements, and maps them to the same classes that were used by Gavriluţ and Pop~\cite{gavrilut_traffic-type_2020}.
Similarly, this method requires that the network operator manually configures the network based on stream assignment.

The \acf{IIC} released a whitepaper~\cite{iic} on the traffic classes found in typical industrial networks.
This can also be used to map \ac{TSN} streams to one of nine priority classes.
These classes range from isochronous streams with hard real-time requirements to best-effort traffic.
Furthermore, they define the required \ac{QoS} guarantees and TSN mechanisms for each class.

In contrast to existing work, our proposed architecture automatically integrates eligible non-TSN traffic into TSN networks and leverages the existing signaling and configuration mechanisms of \ac{TSN}.
The proposed mechanism does not rely on manual configuration by the network operator.

\subsection{Periodicity Detection}

We review related work on periodicity detection.
With periodicity detection, a time series of repeating data patterns is given, and the appropriate period of the repeating data pattern has to be determined.
Existing approaches are divided into two categories.
In the first category, spectral analysis and periodograms are used. 
In the second category, the data is divided into segments of equal length.

\subsubsection{Spectral Analysis and Periodograms}
Vlachos et al.~\cite{vlachos_periodicity_2005} propose the AUTOPERIOD method as a periodic analysis method.
It first uses a \acf{FT} periodigram to select period candidates.
Then, the selection is refined via the \acf{ACF} to determine a single period.
Puecht et al.~\cite{lemaire_fully_2020} improve the AUTOPERIOD method for noisy data.
They leverage clustering algorithms on the selected period candidates and apply a low-pass filter to remove artifacts.
Another improvement by Wen et al.~\cite{wen2021robustperiod} is designed to detect multiple periods in the dataset.
Their presented method results in an increased F1-score compared to AUTOPERIOD on selected datasets.
A similar approach by Shehu and Harper~\cite{ShHa23} uses the Lomb-Scargle method instead of the \ac{FT} to handle unevenly spaced input data.

\subsubsection{Segment Division}
Yuan et al.~\cite{yuan2017periodicities} do not rely on spectral analysis. 
They divide the time series into time slots identified by a period length and offset.
Each timeslot is then scored based on the events covered in each instance of the timeslot.
The resulting output of the algorithm is one or more time slots that maximize this score.
Similarly, Li et al.~\cite{li_mining_2012} also divide the time series into segments of a set length.
They argue that if a segment length describes the period, then the events are located in a similar part of each segment.
Eslahi et al.~\cite{eslahi_periodicity_2015} also use time slots to detect periodic behavior in order to identify HTTP botnets.
However, their approach requires a specified timeslot length which greatly influences the algorithm output.

\par \medskip 

The periodicity approach presented in this work differs from the summarized related work.
It employs a two-step procedure to determine the period of a time series.
In the first step, it uses a \acf{RNN} to determine whether a time series can be considered periodic.
This provides multiple advantages for the application in industrial networks.
The \ac{RNN} can be trained on data observed in the real network.
This allows the \ac{RNN} to leverage a more flexible definition of periodicity tailored to specific industrial needs.
In the second step, we leverage a heuristic to extract the period of an observed periodic data transmission. 
The periodicity detection is described in \sect{periodicity} and the period extraction in \sect{traffic_description}.

%% file: chapters/neural_networks.tex
\section{Neural Networks}\label{sec:neuralnetworks}
We first introduce the concept of \acfp{NN} and their internal structure.
We then elaborate on how these networks are trained to classify data.
Finally, we introduce \acfp{RNN} and explain their use-case.

\subsection{Feedforward Neural Networks}
\acp{NN} are a machine-learning method inspired by the human brain that can learn hidden patterns and characteristics from data sets.
They consist of interconnected nodes, called neurons, that transform a set of input values into a set of output values.
In a simple feedforward \ac{NN}, the neurons are separated into three layers as illustrated in \fig{FIG2.pdf}.
Here, feed-forward means that data is fed through the network layers in one direction.
Every neuron in the hidden layer is connected to every neuron in the input and output layer, but not to another neuron in the same layer.
As a result, the output of a neuron depends on the output of all neurons in the preceding layers.
\figeps[0.7\columnwidth]{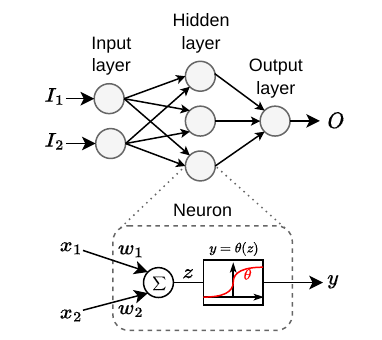}{Architecture of a simple feedforward \ac{NN}. The input data travels through the network in one direction, without any loops. Each Neuron calculates the weighted sum over all inputs and outputs the corresponding value of the activation function $\theta$.}

The transformation inside a single neuron consists of two steps.
First, it calculates the weighted sum over all input values $\sum_i x_i \cdot w_i$.
Here, $x_i$ are the outputs from the previous layers and $w_i$ is the associated weight per input.
Both can be any real number.
Then, the neuron uses a so-called activation function to map the weighted sum to an output value.
The activation function transforms the neuron output and is chosen depending on the application.
An example activation function is the \acf{ReLU}~\cite{nair_rectified_2010} which is $\phi(x) = max(0, x)$.

\subsection{Supervised Training}
Supervised training describes a process to prepare a \ac{NN} before its usage.
It requires a data set of example input and expected output values that represent the actual application data.
For example, a training set for traffic classification could consist of recorded packets and which type of application it belongs to.
During training, the \ac{NN} learns the characteristics and patterns that define the traffic classification.
The trained \ac{NN} can then classify observed network packets.

\subsubsection{Training Process}
The goal of the training process is to adjust the weights of all neurons such that an error function is minimized.
Here, the error function measures the difference between the output of the \ac{NN} and the expected output per sample from the training dataset.
During training, the input samples from the training data set are fed through the network.
Then, the gradient of the error function over the space of neuron weights is calculated, e.g., via stochastic gradient descent.
Based on the gradient, the neurons' weights are adjusted to reduce the error.
One such iteration over the entire training dataset is called an epoch.
Training an \ac{NN} takes multiple epochs until the error value cannot be decreased significantly anymore.

\subsubsection{Overfitting}
An overfitted \ac{NN} performs very well on the training dataset, but not during real-world application.
The reason for overfitting is that the network learns characteristics that are only present in the training dataset.
An overfitted \ac{NN} is only able to detect exact matches of learned patterns while a well-trained \ac{NN} also detects variations of those patterns.
A common approach to counter this effect is to split the training dataset into a training and validation set.
The validation set is not part of the training itself and is used to monitor the overfitting error.
An increase in the error for the validation set after multiple epochs indicates overfitting.
Another approach is neuron dropout where the output of randomly selected neurons is set to zero for an epoch and they are not included in the error gradient calculation.

\subsection{\acfp{RNN}}
\acp{RNN} are specialized \acp{NN} for sequential data, e.g., time series data.
They can detect correlations between consecutive inputs into the neural network.
In essence, the output of a \ac{RNN} does not only depend on the current input but also on previous ones.
Their structure is the same as standard \acp{NN} but the \ac{RNN} neurons store an additional hidden state, the so-called memory.
It is updated with each new input into the neuron and thus also influences any future output.
Currently, the most common architecture for \ac{RNN} neurons are \acf{LSTM}~\cite{hochreiter_long_1997} cells.

%% file: chapters/concept.tex
\section{Integration of \ac{TSN}-unaware Streams}
\label{sec:concept}

In this section, we first suggest a proxy-based integration concept for eligible \ac{TSN}-unaware streams.
Then, we elaborate on the processing steps necessary to integrate \ac{TSN}-unaware streams.

\subsection{Proxy-Based Integration Concept}
We propose an architecture that automates the integration of non-TSN traffic via an additional network entity.
The so-called \acf{SCIP} performs the stream announcement on behalf of \ac{TSN}-unaware applications.
This concept allows \ac{TSN}-unaware streams to benefit from \ac{TSN} features without any changes to the end stations.
\fig{FIG3.pdf} shows an adaption of the fully centralized configuration model that illustrates the novel architecture.
First, the \ac{SCIP} monitors the network traffic for eligible \ac{TSN}-unaware streams.
Once it identifies a new flow, e.g. a real-time video stream, it transmits a stream announcement to the \ac{CNC} \circled{1}.
The CNC then performs the stream admission control process.
It calculates the required network configuration, configures the TSN network components \circled{2}, and sends the device configuration for the stream's end stations back to the \ac{SCIP} \circled{3}.
This configuration message contains the \ac{VLAN} header that the talker has to set in every transmitted packet.
Then, the \ac{SCIP} configures the adjacent switch \circled{4} to add the received \ac{VLAN} tag to every packet of the \ac{TSN}-unaware stream.
More specifically, it configures the IEEE Std. 802.1CB \cite{8021cb} ''IP'' and ''Active Destination MAC and VLAN'' Stream identification functions.

\figeps[\columnwidth]{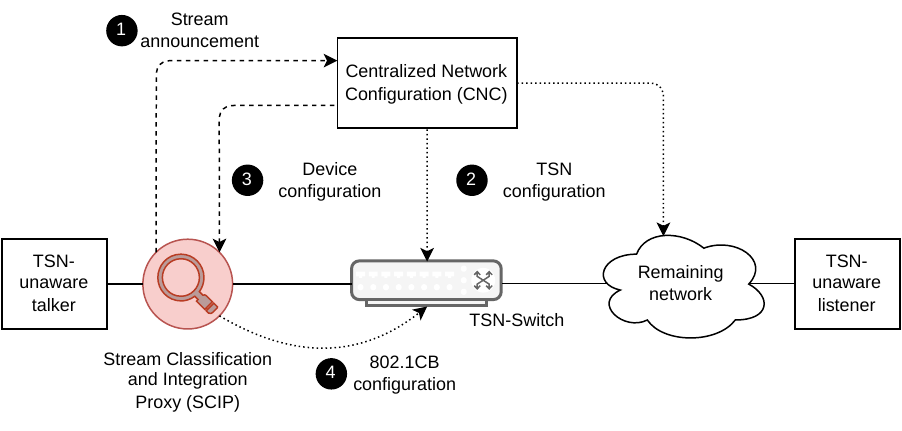}{Extended configuration model in accordance with the fully centralized model. The \acf{SCIP} examines network traffic and invokes stream reservation on behalf of applications. Further, it configures the first bridge on the path to apply 802.1CB stream identification methods.}

\subsection{Stream Processing Stages}
The processing of a new stream in the \ac{SCIP} consists of four stages as visualized in \fig{FIG4.pdf}.
At the beginning of the processing, the \ac{SCIP} has to recognize \ac{TSN}-unaware streams from all observed network packets.
The \ac{SCIP} also records all packets associated with a \ac{TSN}-unaware stream for the next stages.
When the \ac{SCIP} has collected enough information about a stream, it classifies the stream as periodic or non-periodic in the periodicity detection stage (see \sect{periodicity}).
Non-periodic streams are regarded as unfit for integration, and the stream is not processed further, i.e., it is treated as regular \ac{BE} traffic.
If the stream is periodic, the \ac{SCIP} extracts the \ac{TSN} traffic description (see \sect{traffic_description}).
In the last stage, the \ac{SCIP} classifies the stream based on the extracted information to determine the \ac{QoS} requirements (see \sect{qos_classification}).

\figeps[\columnwidth]{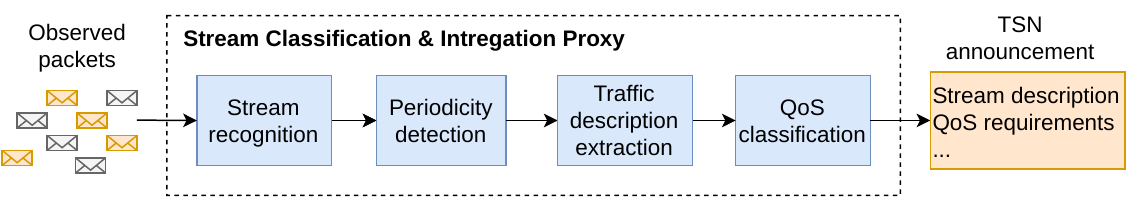}{Processing stages within the \ac{SCIP} for a \ac{TSN}-unaware stream. The input for the processing pipeline are recorded network packets and the output are the parameters required for a \ac{TSN} stream announcement.}

\subsubsection{Stream Recognition}
The integration process starts with the recognition of a new stream in the network.
This means that the \ac{SCIP} receives a packet that does not belong to any known stream.
A packet is associated with a stream based on the IP stream identification function, as defined in the IEEE Std. 802.1CB~\cite{8021cb}.
The \ac{SCIP} then monitors this new stream and records the packets for processing in the following stages.
Streams that are already registered within the \ac{TSN} network have to be filtered out at this stage.

\subsubsection{Periodicity Detection}
The periodicity detection stage determines whether a stream can be integrated into the \ac{TSN} network.
It assumes that a stream has to be periodic in order to have real-time characteristics.
We propose a machine-learning approach using a \acf{DRNN} to detect periodicity in the packet arrival times of a stream.
The biggest advantage of a machine-learning approach lies in its ability to learn from recorded network traffic.
We present the periodicity classification \ac{DRNN} in \sect{periodicity}.
\subsubsection{Traffic Description Extraction}

TSN uses three parameters to describe the bandwidth used by a stream as defined in IEEE Std. 802.1Qcc~\cite{8021qcc}: \textit{Interval (w)}, \textit{MaxFrameSize ($f_{size}$)}, and \textit{MaxFramesPerInterval (m)}.
The parameter \textit{Interval} specifies the length of a sliding time window for devices that do not synchronize their internal clocks with the network.
In addition to that, it is also the maximum delay that a packet should experience in order to arrive before the next transmission cycle starts~\cite{iic}.
At any point in time, the interval window can contain at most \textit{MaxFramesPerInterval} packets.
The third parameter \textit{MaxFrameSize} defines the maximum size of a frame transmitted by the talker.
For standard \ac{TSN} applications, these parameters are selected manually.
We define an algorithm to automatically detect the \textit{Interval} and \textit{MaxFramesPerItnerval} parameters based on packet arrivals in \sect{traffic_description}.

\subsubsection{QoS Classification}
The \ac{TSN} stream announcement contains the stream's \ac{QoS} requirements.
Therefore, a traffic classification approach is used to determine these requirements for an observed \ac{TSN}-unaware stream.
More specifically, the \ac{QoS} requirements are derived from the traffic classification output 
since the \ac{QoS} requirements are known for most applications.
In this work, the Rhebo Industrial Protector \cite{rhebo_protector} is used to discover the applications of recorded streams.
The \ac{QoS} classification is further elaborated on in \sect{qos_classification}.

%% file: chapters/periodicity.tex
\section{Periodicity Detection}\label{sec:periodicity}
In this section, we describe the mechanism to detect whether a TSN-unaware stream is periodic.
We first elaborate on the used artificial dataset and the included types of periodic and aperiodic streams.
This includes a description of the dataset generation process.
Then, we derive the input features and present the full \acf{RNN} classifier architecture.
Last, we validate the periodicity detection mechanism with the generated dataset.

\subsection{Dataset}\label{sec:dataset}
Datasets used in related work (e.g., \cite{vlachos_periodicity_2005}, \cite{yuan2017periodicities}) do not contain labeled periodic and aperiodic network streams.
Thus, we leverage an artificially generated dataset\footnote{The dataset and generation scripts can be found on GitHub~\cite{github}} that contains a variety of periodic and aperiodic streams.
Periodic streams follow a periodic pattern, e.g., packets with (almost) fixed \acp{IAT} or a repeating pattern of \acp{IAT}.
Aperiodic streams do not follow a periodic pattern, i.e., they have seemingly random \acp{IAT}.
\foursubfigeps
{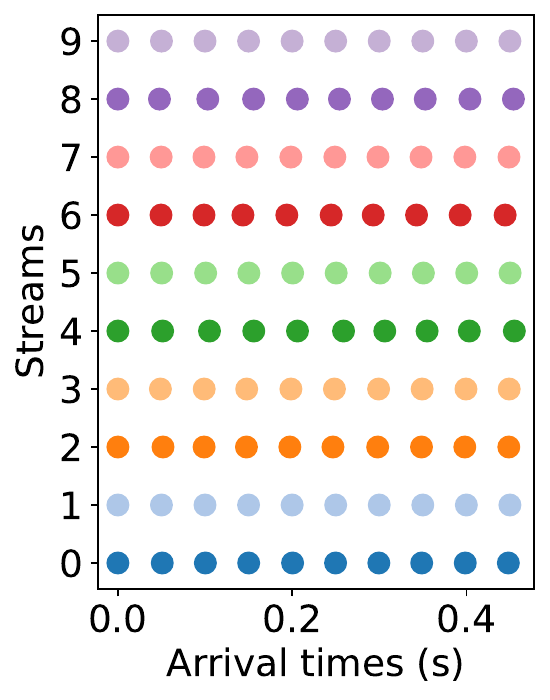}{Pure periodic.}
{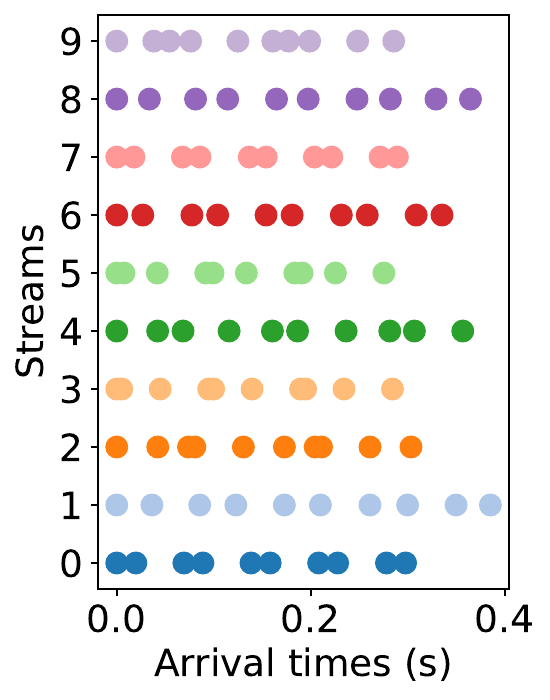}{Periodic pattern.}
{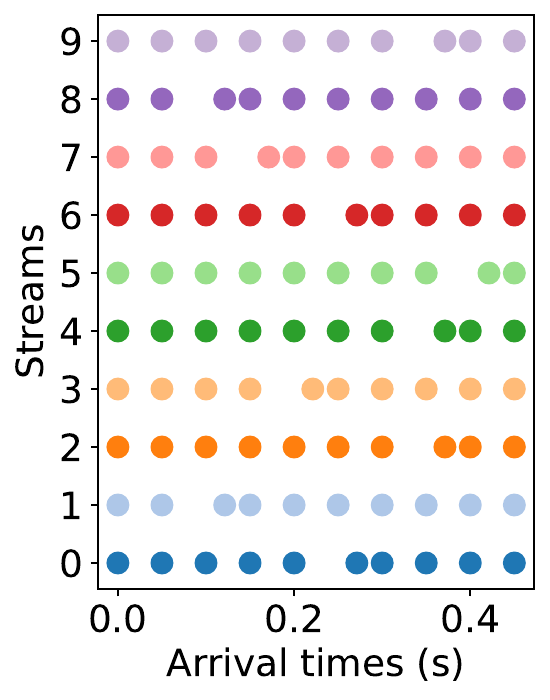}{Near-periodic.}
{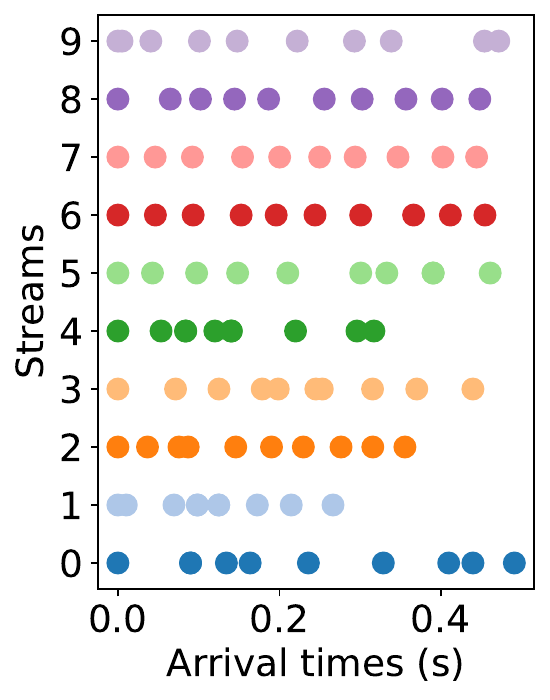}{Aperiodic.}
{The dataset consists of 4 different kinds of streams: aperiodic, near-periodic, periodic, and periodic patterns. Aperiodic and near-periodic streams should be classified as aperiodic. Periodic streams and periodic patterns should be classified as periodic.}

\subsubsection{Stream Types}
Figures \ref{fig:FIG5a.pdf}-\ref{fig:FIG5d.pdf} visualize the stream types present in the dataset.
The x-axis represents the arrival time of a packet, and the y-axis represents different streams.
Pure periodic streams (\fig{FIG5a.pdf}) are streams that have a fixed \ac{IAT} with only little variation.
Streams with periodic patterns (\fig{FIG5b.pdf}) consist of a repeating pattern of \acp{IAT}.
These two types of streams should be classified as periodic.
Near-periodic streams (\fig{FIG5c.pdf} are streams that have a fixed \ac{IAT} but may contain one or more outliers.
Finally, aperiodic streams (\fig{FIG5d.pdf}) are streams with seemingly random \acp{IAT} that do not fit into any of the three other classes.
These two stream types should be classified as aperiodic.

\subsubsection{Generation}
Pure periodic and aperiodic streams with a single packet per period are generated by drawing \acp{IAT} from a normal distribution.
The mean $\mu$ is set to the target period $p$ and the deviation $\sigma = c \cdot \mu$ is derived from the target coefficient of variation c.
Thus, a vector of \acp{IAT} for a stream with $n$ packets is drawn as
\begin{equation}
    \mathbb{X}_n = \{X_0,\dots,X_{n-2}\} \text{ with } X_i {\sim} \max (0, \mathcal{N}(\mu,\sigma^2))
\end{equation}

To produce a periodic pattern, a pattern mask is generated based on the number of packets per pattern $m$.
The mask is a vector of values drawn from a uniform distribution
\begin{equation}
    \mathbb{M}_{m} = \{M_0,\dots,M_{m-2},1\} \text{ with } M_j \overset{\text{i.i.d}}{\sim} \mathcal{U}(0,1)
\end{equation}
We chose the last value in the mask as one to ensure one maximum size IAT.
A stream with a periodically repeating pattern of $m$ packets is generated by applying the mask consecutively to the sampled \acp{IAT}:
\begin{equation}
    \mathbb{X}'_{n}(\mathbb{M}_{m}) = \{X_k \cdot M_{k\ \mathrm{mod}\ m} \text{ }|\text{ } k \in [0..n-2]\}
\end{equation}
Thus, the average \ac{IAT} for a periodic pattern stream with $m$ packets per period is $p/m$.

To generate a near-periodic streams, $n=40$ packets of a pure periodic stream with c = 0.01 are generated.
Then, the arrival time of a single, randomly chosen packet is delayed (not the first and not the last one).
That is, the packet's prededing IAT is extended by a delay $d$, and the packets succeeding IAT is reduced by a delay $d$.
The value $d$ is chosen as large as possible under the condition that the empirical coefficient of variation $\hat{c}$ over all 35 IATs remains smaller than 0.04. Thereby, the stream’s standard deviation over all considered IATs still equals the one of a pure periodic stream.
\tabl{dataset} compiles an overview of the streams in the dataset and the used value ranges for the IAT's coefficient of variation $c$ and for the duration $p$ of the period. 

\begin{tab}{dataset}{Overview of the streams in the artificial dataset and their generation parameters.}
\small
\begin{tabularx}{\columnwidth}{XXXXXXX}\hline
\multirow{2}{*}{Class}             & \multirow{2}{*}{Periodic} & \multicolumn{3}{c}{Periodic pattern}  & \multirow{2}{*}{\shortstack[l]{Near-\\periodic}} & \multirow{2}{*}{Aperiodic} \\
                  &          & $m=2$                 & $m=3$   & $m=4$   &               &           \\\hline
Samples & 2000     & 668                & 666 & 666 & 2000          & 2000      \\
$c_{var}$                 & [0;0.05)      &   [0;0.05)                 &  [0;0.05)   &  [0;0.05)   &      0.01        &      [0.05;1]      \\
$p$                 & $[1\mu s;1s]$     &   $[1\mu s;1s]$                 &  $[1\mu s;1s]$   &  $[1\mu s;1s]$   &      $[1\mu s;1s]$       &      $[1\mu s;1s]$      \\
$n$                 & 36         &   36             &  36 &  36 &    36        &    36     \\\hline
\end{tabularx}
\end{tab}

\subsection{RNN Architecture}\label{sec:rnn_io}
The input feature for the \ac{RNN} is the coefficient of variation $c_n=\frac{\sigma_n}{\mu_n}$ over the recorded \acp{IAT} after $n$ packets of the observed stream.
We use the change in this coefficient over time to differentiate between periodic and non-periodic streams.
The output of the \ac{RNN} is a classification of whether the stream should be regarded as periodic.
It uses the Sigmoid activation function in the last layer that is commonly used for binary classification.
It outputs a real value between zero and one. 
The higher the value, the more confident the network is that the observed stream is periodic.
Between the input and output layers are four hidden layers.
The first and last layers have 200 LSTM cells as neurons each, and the second and third hidden layers have 400 LSTM cells each.
Each neuron in the hidden layers uses the \acf{ReLU} activation function which is commonly chosen for neural networks with multiple hidden layers~\cite{nair_rectified_2010}.
Furthermore, we use dropout factors for the last two layers of 25\% each to reduce the possibility of overfitting.

\subsection{Validation}
We train and validate the \ac{RNN} on the artificial dataset described in \sect{dataset}.
Therefore, we split the dataset into a test and training set of equal size.
The metrics Accuracy, Precision, Recall, and F1 score are used to measure the performance of the \ac{RNN}.
We first determine the number of recorded packets required to accurately classify a stream.
Then, we show the overall accuracy of the \ac{RNN} on the entire dataset.

\subsubsection{Metrics}
Accuracy, Precision, Recall, and F1 score are commonly used for validating machine-learning classification methods.
They are based on the total number of \acf{TP}, \acf{TN}, \acf{FP}, and \acf{FN} classifications.
Here, positive means a classification as periodic, and negative as non-periodic.
Using this, the validation metrics are calculated as
\begin{align}
\text{Accuracy} =& \text{ } \frac{\text{TN} + \text{TP}}{\text{TN} + \text{FP} + \text{TP} + \text{FN}} \\[0.5ex]
\text{Recall} =& \text{ } \frac{\text{TP}}{\text{TP} + \text{FN}} \\[0.5ex]
\text{Precision} =& \text{ } \frac{\text{TP}}{\text{TP} + \text{FP}} \\[0.5ex]
\text{F1-Score} =& \text{ } 2 \cdot \frac{\text{Precision} \cdot \text{Recall}}{\text{Precision} + \text{Recall}}
\end{align}
Accuracy measures the total percentage of correct classifications (either as periodic or non-periodic).
The Recall states the percentage of the detected periodic streams.
Precision is the percentage of correct classifications as periodic.
The F1-Score combines Recall and Precision into a single value via the harmonic mean.
For this work, the Precision is especially important.
With a low Precision, more non-periodic streams are classified as periodic.
Since the integration of non-periodic streams can be harmful to the network, the Precision should be as high as possible.

\subsubsection{Required number of input samples}
We compute the validation metrics by applying the \ac{RNN} on the test split of the dataset.
The \ac{RNN}s real output between zero and one is converted into a binary classification using a threshold value.
If the output is larger than the threshold, the stream is classified as periodic, and aperiodic otherwise.
This means that a higher threshold requires a more confident \ac{RNN} output for a positive classification.

Figures \ref{fig:FIG6a.pdf}-\ref{fig:FIG6d.pdf} visualize Accuracy, Precision, Recall, and F1-Score based on the number of observed packets for each stream in the dataset.
Here, the x-axis describes the number of packets recorded, i.e., the number of inputs for the \ac{RNN} per stream.
The y-axis displays the chosen metric calculated over the \ac{RNN} output for all streams after $x$ packets.
The different lines describe the results for different classification thresholds.
For $x < 9$ the metrics are inaccurate since streams with four packets per pattern cannot be classified correctly.
\foursubfigverticaleps
{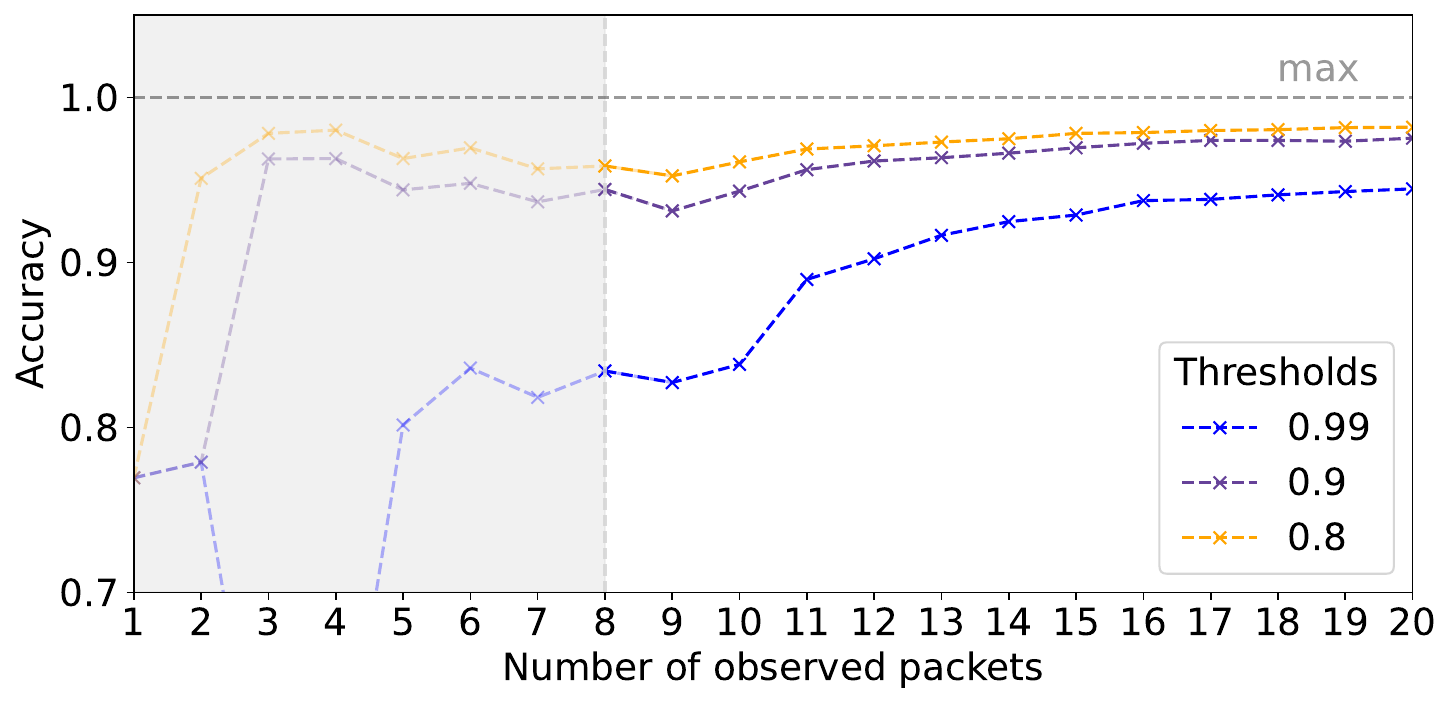}{Accuracy.}
{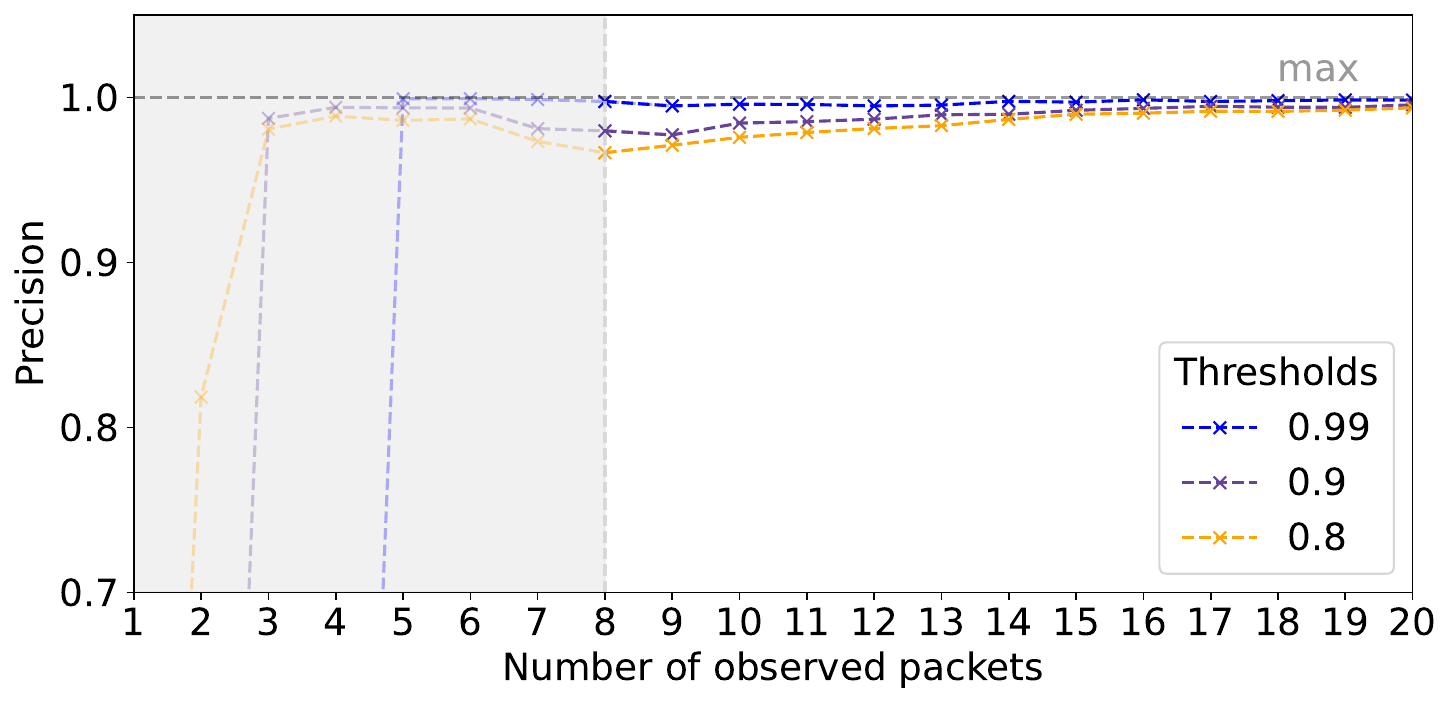}{Precision.}
{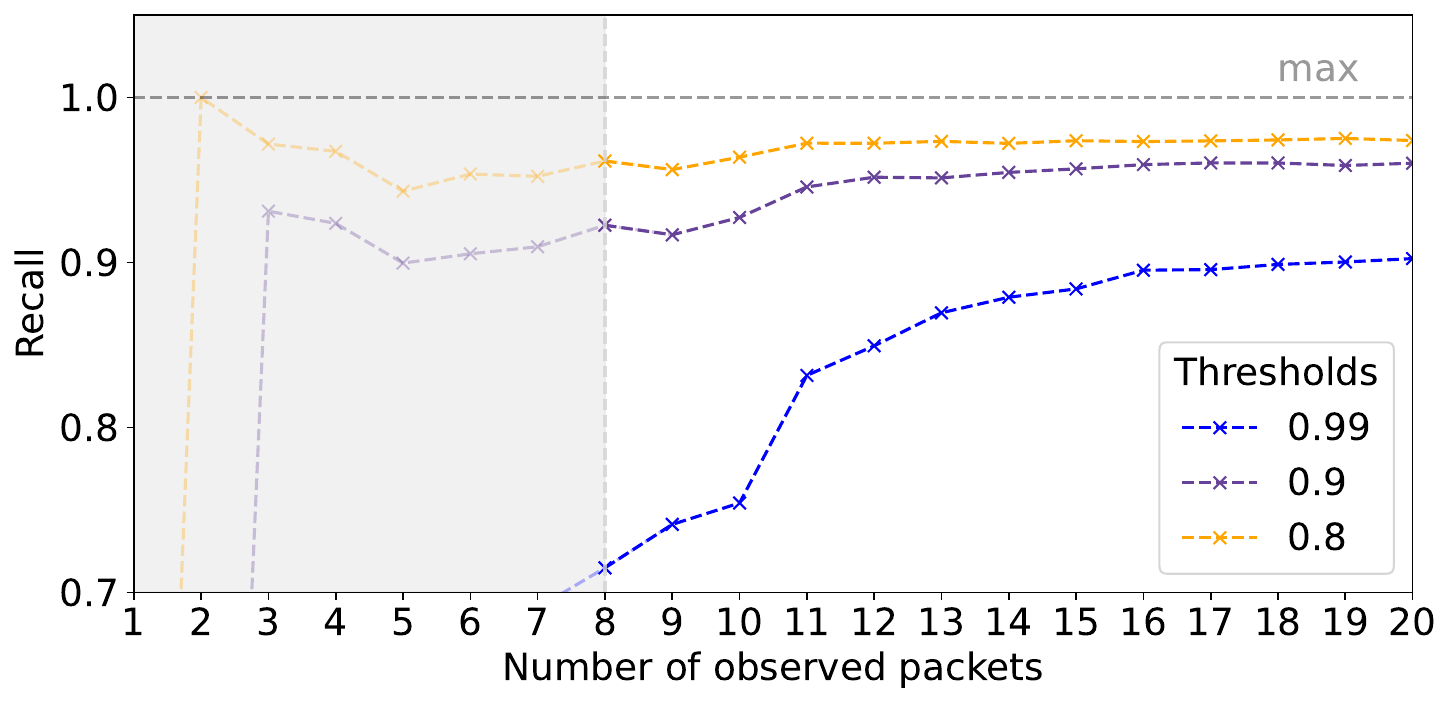}{Recall.}
{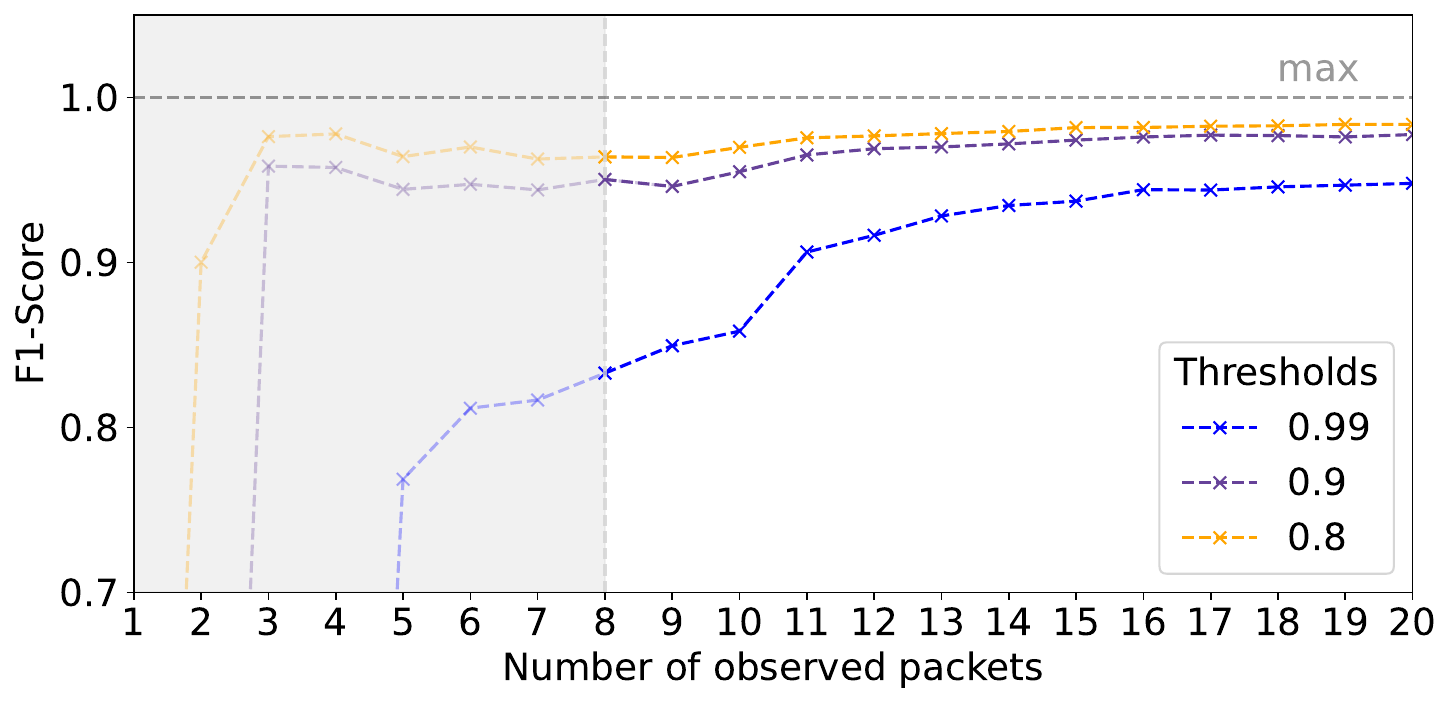}{F1-Score.}
{
Accuracy, Precision, Recall, and F1-Score after x observed packets calculated over all streams from the dataset.
For each stream, the RNN is applied after each new packet, and the results are stored.
Then, the number of TP, FP, TN, FN is derived and used to calculate the F1-Score.
There is a transient phase in the beginning until 8 packets have been recorded.
The color of the lines indicates the confidence value used for the classification.
}
The figures illustrate that all metrics improve for more observed packets (x) and converge toward a stable value after approximately 15 packets.
They also show that a larger threshold increases the Precision but decreases all other metrics.
However, the absolute Precision increase is smaller than the decrease of the other metrics.
Based on the figure, we determine that the \ac{RNN} provides an accurate output after 15 packets.
If a high threshold is chosen, the \ac{RNN} should only be considered after at least 20 packets.

\subsubsection{Overall Performance}
Table~\ref{tab:RNN_results} shows a variation of selected thresholds and the resulting validation metrics after 20 observed packets.
With the highest threshold of $0.99$, the Precision is $99.83\%$ while only $90.38\%$ of the periodic streams are detected.
\begin{tab}{RNN_results}{Results of RNN validation after 20 observed packets based on the used classification threshold.}
\vspace{0.3cm}
\small
    \begin{tabularx}{\columnwidth}{XXXXX}
\hline
\textbf{Threshold} & \textbf{Accuracy} & \textbf{Recall} & \textbf{Precision} & \textbf{F1-Score} \\ \hline
0.99               & 94.57\%           & 90.38\%            & 99.83\%         & 94.87\%            \\ 
0.9                & 97.61\%           & 96.14\%            & 99.53\%         & 97.81\%           \\ 
0.8                & 98.23\%           & 97.42\%            & 99.38\%         & 98.39\%           \\ 
0.5                & 98.76\%           & 98.94\%            & 98.84\%         & 98.87\%           \\
0.3                & 98.68\%           & 99.24\%            & 98.40\%         & 98.81\%           \\ \hline
\end{tabularx}
\end{tab}
Choosing lower thresholds decreases the number of \acp{FN} but also increases the number of \acp{FP}.
As a result, the Accuracy increases to $98.76\%$, and the Recall to $98.94\%$ for a threshold of $0.5$.
While the Precision decreases down to $98.84\%$, the F1-Score is still higher than for the $0.99$ threshold at $98.87\%$.
If the threshold is chosen too low at $0.3$, than the Accuracy and F1-Score begin to decrease again.
This shows that the threshold should be chosen between $0.5$ and $0.99$, based on the severity of \acp{FP}.

%% file: chapters/traffic_description_algorithm.tex
\section{Traffic Description}\label{sec:traffic_description}

In this section, we describe the problem of finding an appropriate traffic description for streams that have been recognized by the periodicity detection.
We present a heuristic to solve this problem and evaluate it on an artificial dataset.

\subsection{Heuristic for Finding a Streams Period}
A traffic descriptor for TSN consists of a 3-tuple $D=(w, m, f_{max})$ and implies that a conforming stream has at most $m$ frames of maximum size $f_{max}$ in any left-side open sliding window of duration $w$. 
For an observed traffic stream with $n$ frame arrivals at time instants $t_i$, $0\leq i<n$, an upper bound for a window size $w$ with at most $m$ frames can be computed by
\begin{equation}
w(m)=min\left(t_{i+m}-t_i | i \in [1, ..., n-m]\right)
\end{equation}
leading to a traffic descriptor $D(m)=(w(m), m, f_{max})$. 
Assuming that any observed traffic stream contains at least two periods, a set of potential traffic descriptors can be derived by
\begin{equation}
\mathcal{D}= \{ D(m) | 1\leq m \leq \floor{\frac{n}{2}}\}
\end{equation}
From this set the most appropriate traffic descriptor needs to be chosen. 
Therefore, we assess how well a traffic descriptor $D(m)$ fits an observed periodic stream. 
To that end, we define $u_m(t)$ which counts the number of packet arrivals in the observed stream within the interval $(t,t+w(m)]$:
\begin{equation}
    u_m(t) = |\{t_i | t < t_i \leq t + w(m) \}|
    \label{eq:utilization} 
\end{equation}
Due to construction, we have $u(t)\leq m$ for all $t_0\leq t\leq t_{n-1}-w(m)$. 
We utilize this observation to assess the suitability of the traffic descriptor $D(m)$. 
To that end, we measure by how many packets $u(t)-m$ the stream deviates on average over time from the proposed periodic pattern, i.e., $m$ packet arrivals within any interval $w(m)$ within the observed intrval $[t_0;t_{n-1}]$:
\begin{equation}
    \delta(m) = \frac{\int_{t_0}^{t_{n-1}-w(m)} m - u_m(t) dt}{t_{n-1}-w(m)-t_0}.
    \label{eq:bacc}
\end{equation}
Finally, the stream descriptor is selected from $\mathcal{D}$ whose number of packets $m^*$ minimizes the deviation:
\begin{equation}
m^*=\underset{m}{\mathrm{argmin}}\left(\delta(m) | 1\leq m \leq \floor{\frac{n}{2}}\right)
    \label{eq:objective}
\end{equation}

\subsection{Evaluation}
We evaluate the heuristic to find a stream's period on the dataset presented in \sect{periodicity}.
We only consider periodic streams and streams with periodic patterns as the traffic description heuristic is only applied to periodic streams.
First, we illustrate the method for different periodic patterns and then we show its outcome on the dataset.

\subsubsection{Illustration of the Method}\label{sec:illustration}

To illustrate the method, the traffic description heuristic is applied to the streams of the dataset.
The periodic streams are groped by the length of their generating pattern $m_g$.
\fig{FIG7a.pdf} to \fig{FIG7d.pdf} visualize the distribution of $\delta(m)$ for each group as boxplots.
The x-axis indicates all possible values for $m_g$ and the y-axis shows the corresponding value for $\delta(m)$.
The boxes show the lower and upper quartiles of $\delta (m)$ with the median marked as a horizontal red line.
Whiskers mark the minimum and maximum value of $\delta(m)$.
As a visual aid, the boxes and whiskers where $m = m_g$ are colored blue and marked with an additional dashed vertical line.
\foursubfigeps
{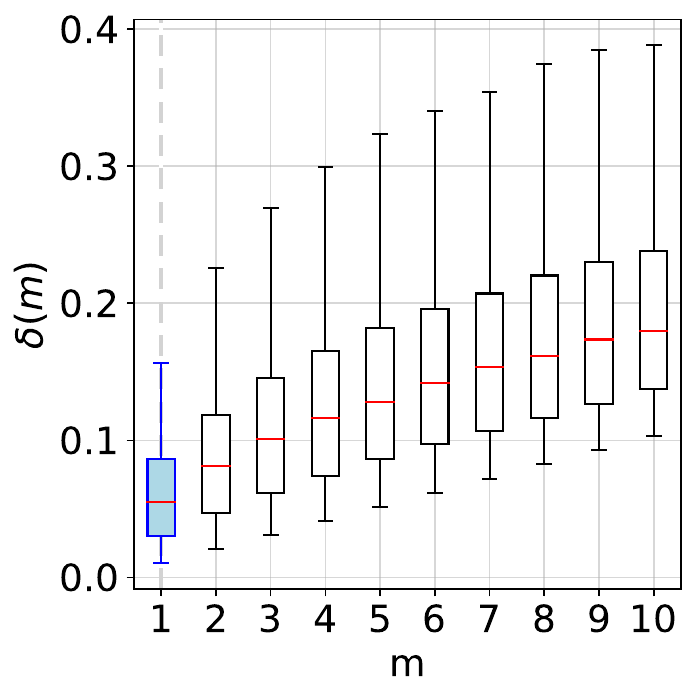}
{One-packet period ($m_g = 1$).}
{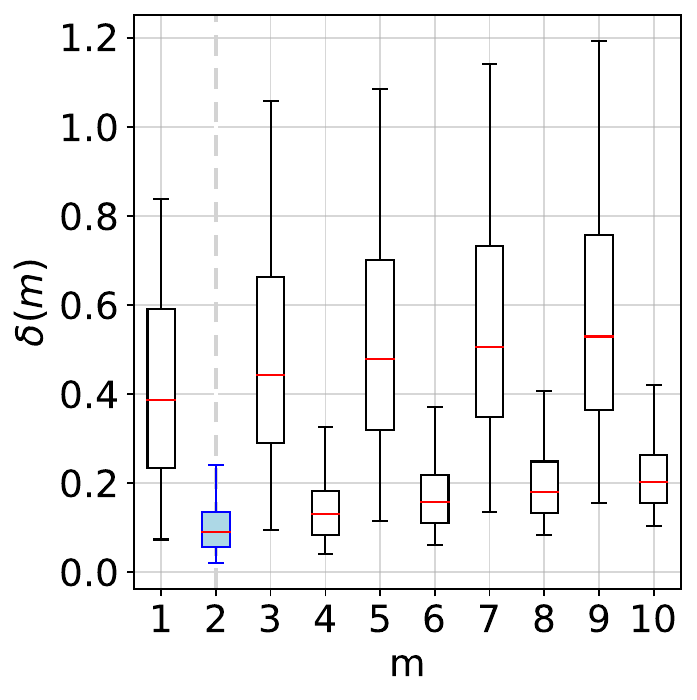}
{Two-packet period ($m_g = 2$).}
{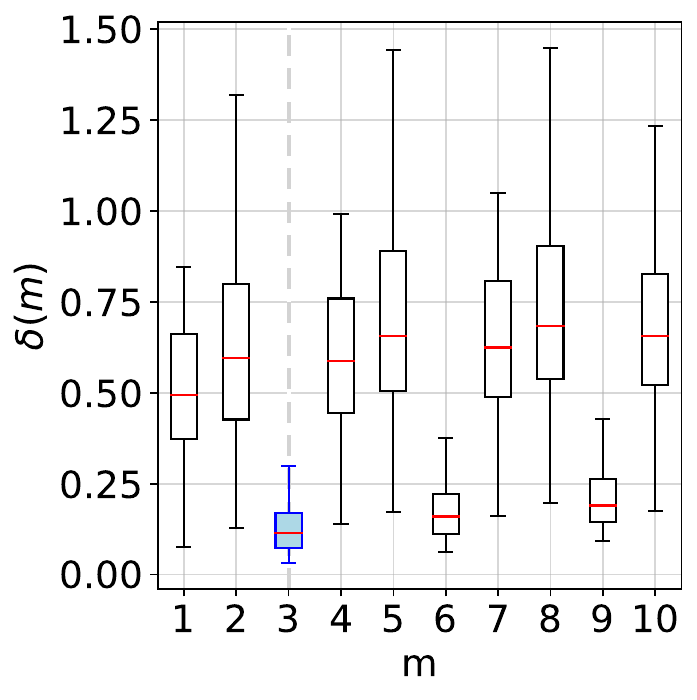}
{Three-packet period ($m_g = 3$).}
{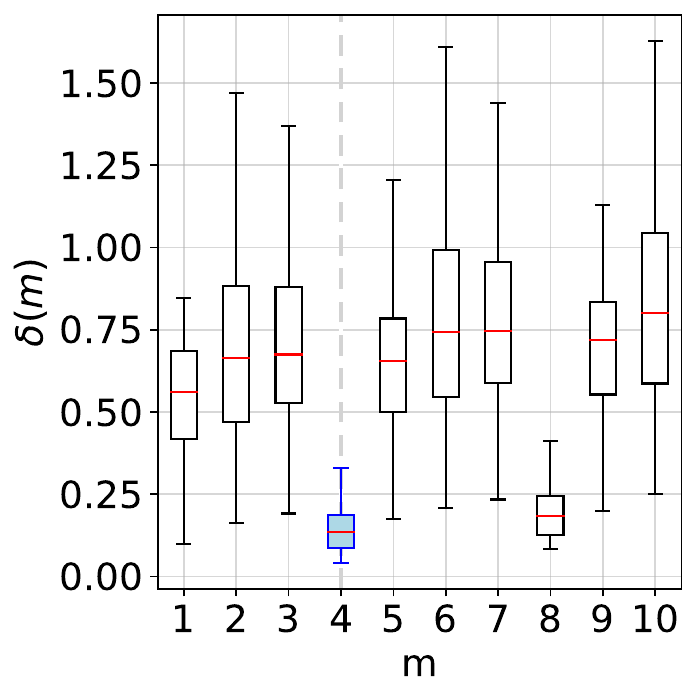}
{Four-packet period ($m_g = 1$).}
{Boxplots of the deviation $\delta(m)$ for different periodic streams from the artificial dataset~\cite{github}.
Each column belongs to a set of periodic streams with $m_g$ packets per period in the real traffic description.
The number of packets per period increases by one from left to right.
The boxplot for the $m^*$ that matches $m_g$ is marked in blue and with a dashed vertical line.}

The figures show that all considered measures like median etc. are lowest for $m=m_g$.
That means, choosing $m^*$ as the lowest $m$ most probably leads to the number of packets $m_g$ for which the periodic pattern was generated. 
The next lowest values of $m$ are multiples of $m_g$. 
That means if $\delta(m_g)$ is by statistical variations now lowest, $k\cdot m_g$, $k\in \mathbb{N}$, will most likely be the lowest value.

\subsubsection{Correctness}
We evaluate the correctness of the traffic description by comparing the outputs $m^*$ to the real traffic description $m_g$ for each stream in the dataset.
Similar to \sect{illustration}, the results are grouped according to size of the generating periodic pattern $m_g$.
\tabl{TD_results} shows how often traffic description heuristic found different values for $m^*$  for different values of $m_g$.
The values $m^*$ are printed bold if they correspond to the size of the generating pattern of the traffic stream.
The resulting percentage of correct descriptions is listet in the last row.
The results show that the heuristic found the correct period size in most cases, i.e., $m^*=m_g$.
The dataset contains 4000 streams and for 3894 streams the period was correctly identified.
This corresponds to a high overall accuracy of 98.00\%.
In cases where $m^*\neq m_g$, the heuristic extracts a multiple of the generating $m_g$ or one.
This can indeed happen because the samples are generated randomly so that packets in a pattern of 4 packets are distributed that they effectively produce a pattern of 2 packets or 1 packet.
Therefore, a stream does not match the chosen generation parameters with a small probability.
Thus, the heuristic extracts a traffic description that matches the stream better than the generating parameters.

\begin{tab}{TD_results}{
Evaluation results for the traffic description heuristic over all streams in the dataset.
The columns represent the generating packets per pattern $m_g$ and the rows the output of the heuristic $m^*$.
The cells indicate how often the heuristic output was $m^*$ when the generating pattern contained $m_g$ packets.}
\small   
\begin{tabularx}{\columnwidth}{XXXXX}\hline
$m^*$   & $m_g$=1    & $m_g$=2    & $m_g$=3    & $m_g$=4   \\ \hline
1    & \textbf{1983}     &  9   &   8  &  3  \\
2  &  16   &  \textbf{654}   &  0   & 1  \\
3  &  1  &  0   &  \textbf{645}   & 0   \\
4  &  0   &  5  &  0  & \textbf{653}   \\
5  & 0    &   0  &   0  & 0  \\ 
6 &   0  &  0   &  12  & 0 \\
7  &   0  &  0   & 0   &  0 \\
8  &  0   &  0   &  0   & 8  \\
9  &  0   &   0  &    1 & 0  \\ 
10  &  0   &   0  &  0   &   0 \\
11  &  0   &   0  &  0  & 0 \\
12  &  0    &  0   &  0   &  1\\ \hline
Total & 2000 & 668 & 666 & 666 \\ 
\mbox{$m^*=m_g$} & 99.15\% & 97.90\% & 96.85\% & 98.05\% \\\hline
\end{tabularx}
\end{tab}


%% file: chapters/qos_classification.tex
\section{QoS Classification with Rhebo Industrial Protector}
\label{sec:qos_classification}

In this section, we explain an approach to determine the \ac{QoS} requirements of a periodic \ac{TSN}-unaware stream.
First, we summarize how the Rhebo Industrial Protector is used to detect the application of an observed stream.
Then, we explain how the detected application can be used to determine the stream's \ac{QoS} requirements.

\subsection{Application Detection}

The Rhebo Industrial Protector is a specialized network intrusion detection system designed for monitoring \acf{ICS}. 
It is connected to switches within the ICS to gather and scrutinize network traffic for anomalies.
The switches in the ICS are configured to mirror traffic to the Rhebo Industrial Protector. 
The \acf{DPI} engine of the Rhebo Industrial Protector collects all network traffic and categorizes it into so-called conversations.
A conversation represents a type of communication between two hosts, including source and destination MAC/IP addresses, protocols and ports, \ac{VLAN} IDs, protocol functions, and throughput.
For each conversation, the \acf{DPI} engine detects the application and forwards it to the \ac{SCIP} for \ac{QoS} classification.

\subsection{QoS Parameter Extraction}

For a \ac{TSN} stream announcement, the following \ac{QoS} parameters are required: stream rank, maximum latency, and the required number of redundant disjoint paths.
The \ac{SCIP} uses an internal database to derive the \ac{QoS} parameters of a stream from the detected application.
This is possible since the \ac{QoS} requirements of most applications, e.g. \ac{VoIP}, are already known.
If the application is unknown, then the extracted period is used to choose one of the traffic classes proposed by the \ac{IIC}~\cite{iic}.
The \ac{QoS} parameters are then derived from the traffic class.

%% file: chapters/evaluation.tex
\section{Experimental Evaluation}\label{sec:evaluation}
This section describes the functional evaluation of the autonomous integration concept from \sect{concept} through a network simulation.
We use the network simulator OMNeT++ 6.0.1~\cite{noauthor_omnet_nodate} and the simulation library INET 4.4~\cite{noauthor_inet_nodate}.
First, we explain the design of the simulated network.
Then, we describe the methodology and the performed experiment.

\subsection{Network Design}

The goal of this evaluation is to implement and simulate the stream integration via the \ac{SCIP} in a TSN network.
We also analyze the impact that the concept has on the experienced \ac{QoS} of a TSN-unaware stream.
\fig{FIG8.pdf} illustrates the topology of the simulated network used in this evaluation.
\figeps[\columnwidth]{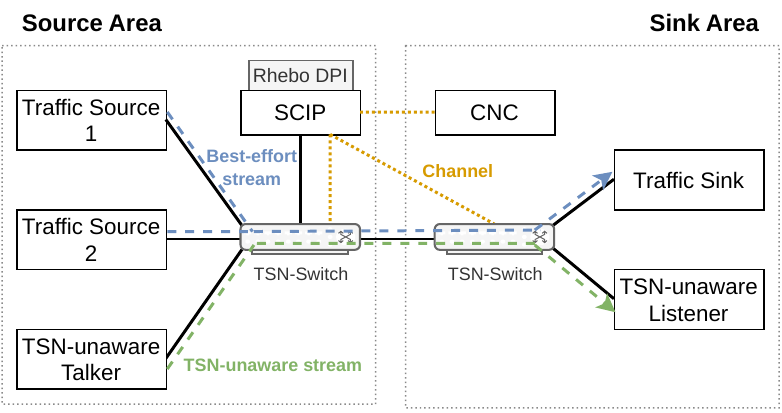}{Architecture of the simulated network used for the experimental evaluation. The link between the two areas serves as the bottleneck to simulate an overloaded network. }

The network consists of two TSN switches that are interconnected via a single link.
This layout is also referred to as a dumbbell network.
It leverages two traffic generators in order to generate an artificial load on the network.
This load creates network congestion on the central link connecting both areas.
As a result, other streams in the network experience decreased \ac{QoS} or even packet loss.

The simulated \ac{SCIP} implements the concept described in \sect{concept} and is connected to a mirroring switch port in the source area.
This mirroring port only clones packets received from the \ac{TSN}-unaware talker.
Thus, the packets arriving at the \ac{SCIP} do not experience any queuing delay such that the inter-arrival times of the packets are not biased.
The Rhebo Industrial Protector's \ac{DPI} engine is directly connected to the \ac{SCIP} for the traffic application classification.

Since no OMNeT++ library implements a \ac{CNC} or \ac{TSN} signaling protocol, we implemented a dummy \ac{CNC} that is directly connected to the \ac{SCIP}.
The \ac{CNC} mimics the behavior of a real \ac{CNC} when replying to stream announcements.
The interface between \ac{CNC} and \ac{SCIP} is modeled after the definitions in IEEE Draft P802.11Qdj~\cite{8021qdj}.
Similarly, the \ac{TSN} switches are configured via a direct channel since they do not support a configuration protocol.

\subsection{Methodology}
We measure the experienced \ac{QoS} of the \ac{TSN}-unaware stream based on the per-packet latency.
It is calculated as the difference between the transmission and reception timestamp.
We then derive the delay and jitter from the per-packet latency.
Furthermore, we measure packet loss based on the number of transmitted and received packets. 

\subsection{Experiment}
In the functional validation, we monitor the end-to-end delay of a TSN-unaware \acf{VoIP} stream in the network described in the previous section.
The TSN-unaware \ac{VoIP} talker transmits a 94 byte packet every 20 ms to the listener.
Two traffic generators increase the load on the network by transmitting 150 ms bursts of 1 Gbit/s every 200 ms.
All links are configured to transmit at 1 Gbit/s which means that the central link connecting both areas experiences a 200\% load.
We refer to the time in which this background traffic is sent through the network as an overload phase.
After an overload phase, the queue emptying phase is the time until the switch queues are cleared.
The switches are configured to have two output queues with strict priority scheduling.
The first queue handles TSN traffic and has a higher priority than the second one.

\fig{FIG9.pdf} shows the measured per-packet latency.
Prior to the overload phase, the end-to-end delay of the \ac{VoIP} stream is almost zero since there are only two switches between source and sink.
In the first load phase, this end-to-end delay increases up to 150 ms.
The jitter reaches a maximum of 20 ms between two packets.
During the queue emptying phase, the \ac{VoIP} packets pile up at the end of the queue.
This causes 7 packets to arrive at the \ac{TSN}-unaware  istener almost simultaneously at $\sim$ 390 ms.
However, no packet loss is experienced since the queues are never filled completely.
\figeps[0.8\columnwidth]{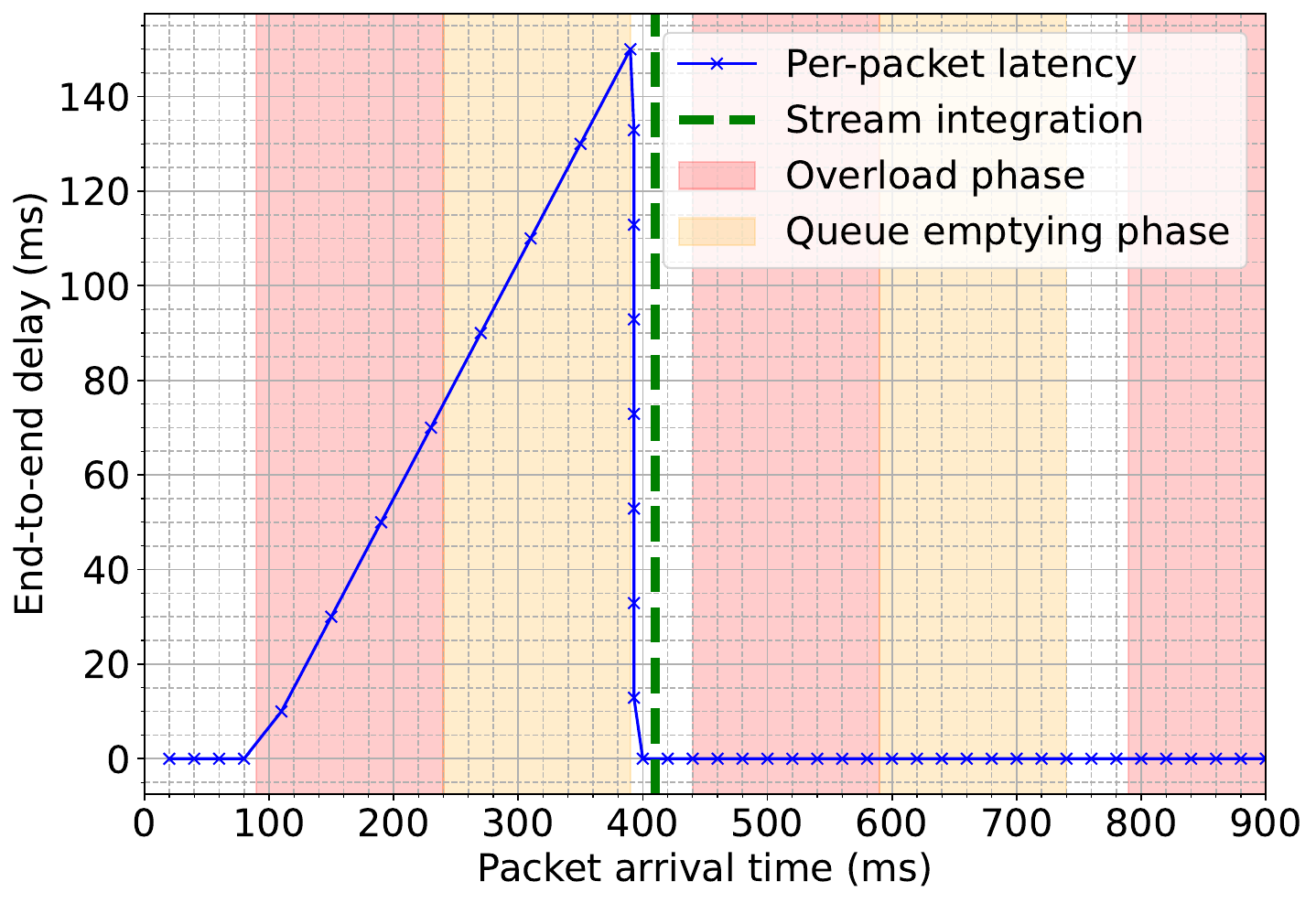}{
Packet delay of a TSN-unaware stream before and after TSN integration. In the first overload and queue emptying phase the stream is not yet TSN-integrated and its packets experience long delays. After TSN-integration at time 400 ms, their delays remain short in spite of overload.
}

After 20 recorded packets, at $\sim$ 400 ms, the \ac{SCIP} announces the stream to the \ac{CNC}.
During the following overload phase, the end-to-end delay does not increase.
This happens because the packets of the \ac{TSN}-unaware stream are inserted into \ac{TSN}'s priority queue on each switch.
As all background traffic is inserted into the default queue, the packets of the \ac{TSN}-unaware stream in the priority queue are not delayed by \ac{BE} traffic any more.

%% file: chapters/discussion.tex
\section{Compatbility with TSN Shaping and Policing}\label{sec:discussion}
\ac{TSN} offers different traffic shaping and policing mechanisms. 
The \ac{CNC} chooses which of these mechanisms are applied to which streams in the network.
This could lead to problems if a mechanism that is incompatible with \ac{TSN}-unaware streams is assigned to such a stream.
\tabl{compatibility} contains a list of the common shaping and policing mechanisms and their compatibility.

\begin{tab}{compatibility}{Compatibility of \ac{TSN} policing and shaping mechanisms with \ac{TSN}-unaware streams.}
\vspace{0.3cm}
    \begin{tabularx}{\columnwidth}{Xcc}
\hline
\textbf{Mechanism} & \textbf{IEEE Standard} & \textbf{Compatible} \\ \hline
\acl{TAS}              & 802.1Qbv~\cite{8021qbv}            & \ding{55} \\ 
\acl{CBS}              & 802.1Qav~\cite{8021qav}            & \ding{51} \\ 
\acl{CQF}              & 802.1Qch~\cite{8021qch}            & \ding{51} \\ 
\acl{ATS}              & 802.1Qcr~\cite{8021qcr}            & \ding{51} \\
\acl{PSFP}             & 802.1Qdj~\cite{8021qdj}            & \ding{51} \\ \hline
\end{tabularx}
\end{tab}

For streams with precise timing requirements, traffic scheduling supported by the \acf{TAS}~\cite{8021qbv} is used.
It requires time synchronization from all involved talkers and thus is incompatible with \ac{TSN}-unaware talkers.
However, the target \ac{TSN}-unaware streams do not require this shaper and its precise guarantees.
The other common \ac{TSN} shapers \acf{CBS}~\cite{8021qav}, \acf{CQF}~\cite{8021qch} and \acf{ATS}~\cite{8021qcr} do not require time synchronization.
As long as the traffic description is accurate, they cope with the integrated \ac{TSN}-unaware streams.

Policing in a \ac{TSN} network is achieved with the \acf{PSFP}~\cite{8021qdj} mechanism.
It is used to enforce an upper rate limit for each stream using \ac{CBS} metering.
If a stream exceeds this rate, \ac{PSFP} may drop single packets or even the entire stream.
As long as the traffic description is accurate, \ac{PSFP} does not interfere with \ac{TSN}-unaware streams.

%% file: chapters/conclusion.tex
\section{Conclusion}
\label{sec:conclusion}

In this paper, we presented an architecture that improves the \ac{QoS} for legacy or other periodic \ac{TSN}-unaware streams with real-time requirements in a \ac{TSN} network.
The so-called \acf{SCIP} detects these streams, extracts the required parameters, and announces them to the network.
It uses a \ac{DRNN} to detect periodic streams and a heuristic to determine an accurate \ac{TSN} traffic description.
The \ac{SCIP} also leverages the Rhebo Industrial Protector to identify the stream's \ac{QoS} requirements.
We evaluated both the DRNN and traffic description heuristic on an artificial dataset.
The \ac{DRNN} achieved an F1-Score of up to 98.87\% and the traffic description detected the correct period for 97.36\% of all streams.
Finally, we evaluated the network architecture in a simulation and showed that the SCIP recognizes a TSN-unaware stream and protects it against overload by TSN integration.
Future work may apply this method to an industrial testbed to automatically protect legacy applications.

%% file: acronyms.tex
\begin{acronym}
\acro{AVB}{Audio Video Bridging}
\acro{CBS}{Credit-Based Shaper}
\acro{CNC}{central network configuration}
\acro{CUC}{central user configuration}
\acro{CQF}{Cyclic Queuing and Forwarding}
\acro{TSN}{Time-Sensitive Networking}
\acro{UNI}{User/Network Interface}
\acro{QoS}{quality of service}
\acro{TAS}{Time-Aware Shaper}
\acro{PCP}{priority code point}
\acro{PLC}{programmable logic controller}
\acro{SCIP}{Stream Classification and Integration Proxy}
\acro{PCP}{Priority Code Point}
\acro{VLAN}{Virtual LAN}
\acro{PSFP}{Per-Stream Filtering and Policing}
\acro{IIC}{Industry IoT Consortium}
\acro{LLDP}{Link Layer Discovery Protocol}
\acro{VoIP}{Voice-over-IP}
\acro{FRER}{Frame Elimination and Replication}
\acro{NN}{Neural Network}
\acro{DRNN}{Deep Recurrent Neural Network}
\acro{RNN}{Recurrent Neural Network}
\acro{LSTM}{Long Short-Term Memory}
\acro{FNN}{Feedforward Neural Network}
\acro{IAT}{inter-arrival time}
\acro{FFT}{Fast-Fourier Transform}
\acro{ACF}{Autocorrelation Function}
\acro{LSSA}{Least-Squares Spectral Analysis}
\acro{ReLU}{Rectified Linear Unit}
\acro{TP}{True Positive}
\acro{TN}{True Negative}
\acro{FP}{False Positive}
\acro{FN}{False Negative}
\acro{FT}{Fourier Transform}
\acro{TT}{Time-Triggered}
\acro{BE}{best-effort}
\acro{ID}{Identification}
\acro{WebRTC}{Web Real-Time Communication}
\acro{ATS}{Asynchronous Traffic Shaper}
\acro{DPI}{Deep Packet Inspection}
\acro{ICS}{Industrial Control Systems}
\end{acronym}